\documentclass{article}
\usepackage[utf8]{inputenc}

\usepackage{amsfonts,amsmath,amssymb}
\usepackage{booktabs}
\usepackage[ruled,vlined]{algorithm2e}
\usepackage{amsmath}
\usepackage{hyperref}
\usepackage{capt-of}
\usepackage{subcaption}
\usepackage{comment}
\usepackage{booktabs}
\usepackage{verbatim}
\usepackage{makecell}
\usepackage{graphicx}
\usepackage{subcaption}
\usepackage{multirow}
\usepackage{algpseudocode}
\usepackage{geometry}
\usepackage{adjustbox}
\usepackage{changepage} 
\usepackage[hypcap=false]{caption}

\begin{document}

{\Large
\textbf\newline{Scaling pattern mining through non-overlapping variable partitioning} 
}
\newline
\\
Leonardo Alexandre\textsuperscript{1,2,3,*},
Rafael S. Costa\textsuperscript{3} and
Rui Henriques\textsuperscript{1,2}
\\

\bigskip
\noindent\textbf{1} INESC-ID, Lisboa, Portugal
\\
\textbf{2} Instituto Superior T\'{e}cnico, Universidade de Lisboa, Lisboa, Portugal
\\
\textbf{3} LAQV-REQUIMTE, Department of Chemistry, NOVA School of Science and Technology, Universidade NOVA de Lisboa, 2829-516 Caparica, Portugal\\
\bigskip

%
%





\noindent * leonardoalexandre@tecnico.ulisboa.pt

\section*{Abstract}
Biclustering algorithms play a central role in the biotechnological and biomedical domains. The knowledge extracted supports the extraction of putative regulatory modules, essential to understanding diseases, aiding therapy research, and advancing biological knowledge. However, given the NP-hard nature of the biclustering task, algorithms with optimality guarantees tend to scale poorly in the presence of high-dimensionality data. To this end, we propose a pipeline for clustering-based vertical partitioning that takes into consideration both parallelization and cross-partition pattern merging needs. Given a specific type of pattern coherence, these clusters are built based on the likelihood that variables form those patterns. Subsequently, the extracted patterns per cluster are then merged together into a final set of closed patterns. This approach is evaluated using five published datasets. Results show that in some of the tested data, execution times yield statistically significant improvements when variables are clustered together based on the likelihood to form specific types of patterns, as opposed to partitions based on dissimilarity or randomness. This work offers a departuring step on the efficiency impact of vertical partitioning criteria along the different stages of pattern mining and biclustering algorithms. \vskip 0.2cm 

\noindent\textbf{Availability:} All the code is freely available at \href{https://github.com/JupitersMight/pattern_merge}{https://github.com/JupitersMight/pattern\_merge} under the MIT license.

\section{Introduction}

Data analysis and machine learning tasks have proven their impact over the years for knowledge extraction \cite{dhillon2019machine, wilson2020survey, alexandre2021mining} and predictive modeling \cite{henriques2014learning}. Biclustering, a well-established algorithm used across multiple domains \cite{naulaerts2015primer, fernando2012effective, glatz2014visualizing, mukherjee2012spotting}, extracts patterns from data. In the biological domain, these patterns can identify functional interactions between biological processes \cite{henriques2015biclustering}, aid in disease understanding \cite{yang2017analysis}, amongst others \cite{xie2019time}. However, biclustering solutions, irrespectively of the problem formulation, are NP-complete. As such, it is crucial that space and time complexity of biclustering algorithms scale in the presence of larger datasets. 
Current approaches can be categorised into three types: i) parallelization principles \cite{xun2016fidoop, chon2018gminer}, where the workload of biclustering can be distributed accross multiple machines (nevertheless susceptible to a huge network overhead) or multi-threaded processing where sequential algorithms are parallelized \cite{liu2007optimization} (resulting in out-of-memory in the presence of large data), ii) approximate searches \cite{djenouri2017combining}, for example defining bio-inspired metaheuristic to produce patterns, filtering out spurious associations and thus reducing overall computational complexity (susceptible to overlooking relevant patterns), and, iii) problem decomposition and partitioning strategies \cite{djenouri2019highly, djenouri2021exploring}, which split the biclustering problem into sub-problems by partitioning the database. Our solution fits into the latter partitioning strategies. Some of the existing solutions logically divide the dataset into a number of non-overlapping horizontal partitions and then generate a set of all potential large itemsets to estimate the support of the pattern reducing the search space \cite{savasere1995efficient}, whilst others make use of MapReduce frameworks to partition the search space evenly among processing units \cite{moens2013frequent}. Vertical state-of-the-art approaches partition a transactional dataset using clustering techniques\cite{djenouri2019highly, djenouri2021exploring}, creating overlapping variable clusters. Their intuition is that by grouping similar observations together they minimize the shared variables between the clusters. 

To the best of our knowledge, current state-of-the-art approaches use classical similarity criteria. As such, we propose a methodology that partitions the dataset into non-overlapping clusters based on variable similarity. This similarity derives from the likelihood that these variables form patterns with a specific coherence \cite{madeira2004biclustering}. Our hypothesis is that by grouping these variables by similarity, and not randomly or based on dissimilarity, the resulting patterns from the clustered variables will be less likely to be further expanded when merging with patterns extracted from other partitions, thus reducing the time complexity of a full run of the algorithm. Given the aforesaid, to correctly partition the dataset we propose a new set of similarity metrics that captures the likelihood of variables forming the desired type of pattern. The gathered results in this study show that, while there are no consistent improvements on the merging execution time with different approaches, on some datasets the execution time with the similarity approach yields statistical significant improvements on the overall execution time. 

The rest of this paper is organized as follows. The Background section formalizes the biclustering task. Methodology further details our proposed solution, the partitioning, merging step, and the experimental setup. Results and Discussion section presents the collected results of the experimental setup as well as an objective view on them. Finally, the major takeaways are highlighted.



\subsection*{Background}

Given a matrix $A$ in the form of $A=(X, Y)$, with a set of observations $X=\{x_1, ..., x_N\}$, variables $Y=\{y_1, ..., y_M\}$, and elements $a_{ij}\in\mathbb{R}$ observed for observation $x_i$ and variable $y_j$, a bicluster $B=(I,J)$ is a $n \times m$ subspace, where $I = (i_1, ..., i_n) \subset Y$ is a subset of variables and $J = (j_1, ..., j_m) \subset X$ is a subset of observations. Multiple biclustering algorithms \cite{madeira2004biclustering, henriques2014bicpam} implement the biclustering task that aims at identifying a set of biclusters $B=(B_1,..,B_k)$, that satisfy specific criteria such as, \textit{homogeneity}, \textit{statistical significance}, and \textit{dissimilarity}.

The aforementioned properties are commonly guaranteed through the use of a merit function, such as the variance of the values in a bicluster \cite{madeira2004biclustering}, guiding the formation of biclusters in greedy, exhaustive, and stochastic/parametric searches determining their coherence, quality and structure (\textit{homogeneity}), statistical significance tests that guarantees that the probability of a bicluster's occurrence (against a null model) deviates from expectations\cite{henriques2018bsig} (\textit{statistical significance}), and additional criteria further placed to guarantee the absence of redundant biclusters (number, shape, and positioning)\cite{henriques2017bicpams}(\textit{dissimilarity}).

\begin{figure}[h!]
  \centering
  \includegraphics[width=0.7\textwidth]{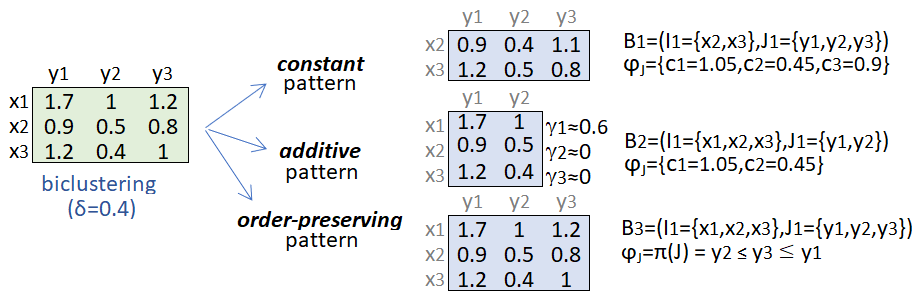}
  \caption{\small Biclustering with varying homogeneity criteria: three biclusters were found under a constant, additive and order-preserving assumption. Illustrating, constant bicluster has pattern (value expectations) \{$c_1=1.05,c_2=0.45,c_3=0.9$\} on $x_2$ and $x_3$ observations, while the order-preserving bicluster satisfies the $y_1\ge y_2\ge y_3$ permutation on \{$x_1,x_2,x_3$\} observations. (adapted from \cite{alexandre2021mining})}
\vskip -0.3cm
  \label{bics_types}
\end{figure}

Given a bicluster $B=(I,J)$, its \textbf{pattern} $\varphi_{J}$ is the set of expected values in the absence of adjustments and noise, as illustrated in Figure \ref{bics_types}. A bicluster \textbf{pattern} is:

\begin{itemize}
    \item constant overall if for all $i \in I$ and $j \in J$, $a_{ij} = \mu+\eta_{ij}$, where $\mu$ is the typical value and $\eta_{ij}$ is the observed noise;
    \item constant on variables, i.e. pattern on observations, if $a_{ij} = \mu_j+\eta_{ij}$, where $\mu_j$ represents the expected value in variable $y_j$;
    \item additive if for all $i \in I$ and $j \in J$, $a_{ij} = \mu_j + \gamma_i+\eta_{ij}$ where $\mu_j$ represents the expected value in variable $y_j$ and $\gamma_i$ the adjustment for observation $x_i$;
    \item multiplicative if for all $i \in I$ and $j \in J$, $a_{ij} = \mu_j \times \gamma_i+\eta_{ij}$ where $\mu_j$ represents the expected value in variable $y_j$ and $\gamma_i$ the adjustment for observation $x_i$;
    \item order-preserving on variables if there is a permutation of $J$ under which the sequence of values in every observation is strictly increasing. Likewise, order-preserving on observations if there is a permutation of $I$ under which the sequence of values in every variables is strictly increasing.
\end{itemize}

The support $\Phi$ of the bicluster \textbf{pattern} $\varphi_{J}$, defined as $\Phi(\varphi_{J})$, is the number of observations containing the bicluster pattern $\varphi_{J}$. The length of the bicluster \textbf{pattern} $\varphi_{J}$, defined as $|\varphi_{J}|$, is the number of variables contained in the bicluster pattern. A maximal bicluster is a bicluster that cannot be extended with additional observations or variables while still satisfying the imposed criteria.

\section{Methodology}

To scale biclustering algorithms in the presence of a high number of variables we propose a decomposition method. We start by vertical partitioning the original dataset into subspaces, containing fewer variables but maintaining the same number of observations. These subspaces are put together by grouping together variables based on their likelihood to create a pattern. Biclustering is then applied to each partition, extracting a set of partition-specific patterns. Finally, these patterns are then merged with each other to guarantee the maximality of the final biclusters. Figure \ref{pipeline} illustrates the proposed solution.

\begin{adjustwidth}{-2.8cm}{0cm}
\vspace{0.5cm}
\begin{minipage}{1.35\textwidth}
\includegraphics[width=\textwidth]{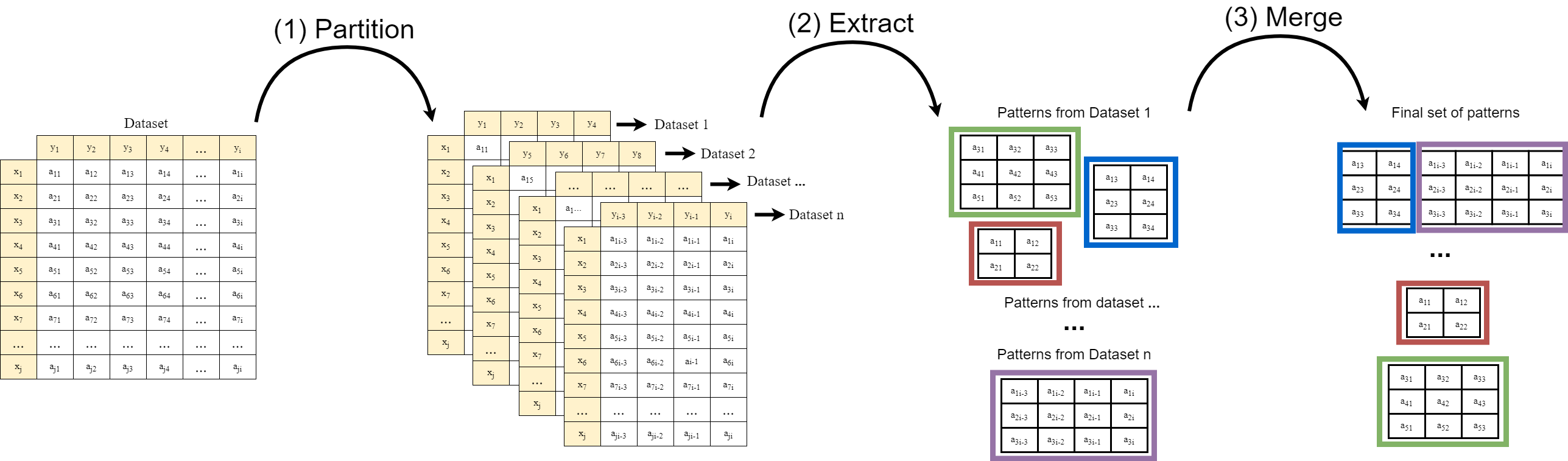}
\captionof{figure}{Overview of the proposed workflow. Starting from left to right, (1) a dataset is partitioned into subspaces containing the original set of observations but only a subset (distinct) of variables grouped by similarity, (2) the application of a biclustering algorithm to extract a set of partition-specific patterns per partition, and (3) the merging of those to form a final set of patterns.}
\label{pipeline}
\end{minipage}
\end{adjustwidth}

\subsection{Partitioning}

To correctly partition the dataset into data subspaces we apply a modified version of agglomerative clustering. Agglomerative clustering iteratively groups together clusters (in the beginning each variable is a cluster) based on a similarity matrix. In this case, we also impose an upper limit on the number of variables per cluster so, whenever a cluster reaches this upper limit, it no longer is considered to form a new cluster. This upper limit ensures that clusters are evenly distributed in the number of variables they contain.

\begin{figure}[!ht]
  \centering
    \includegraphics[width=0.6\textwidth]{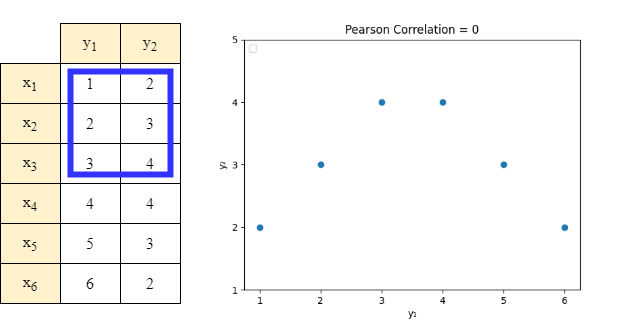}
    \caption{Example of the Pearson correlation metric using two variables, $y_1=[1,2,3,4,5,6]$ and $y_2=[2,3,4,4,3,2]$. In this case the Pearson correlation is equal to zero but there is an additive bicluster pattern with $I=\{y_1,y_2\}$ and $J=\{x_1,x_2,x_3\}$.}
    \label{pearson_example}
\end{figure}

A very important part of this process is that the values in the similarity matrix correctly express the likelihood of variables to form robust patterns with a given coherence. To this end current similarity metrics can be used to capture the likelihood of two variables forming a pattern. An example of said metrics are: 1) Jaccard coefficient \cite{aggarwal2015association}, for patterns exhibiting constant coherence,  2) the Pearson correlation, for patterns exhibiting additive and multiplicative coherence, and, 3) Tau Rank, for patterns exhibiting order-preserving coherence. Although these metrics can be used, they fail to capture relevant cases. Figure \ref{pearson_example} provides an example where the Pearson correlation fails to capture a pattern contained in half of the observations.

For example, the Pearson correlation devalues some cases when dealing with additive and multiplicative patterns. Considering the aforementioned case (Figure 3), Pearson correlation is zero, yet there is an additive pattern with $50\%$ support, $P1 = \{x_1,x_2,x_3\}$.

\begin{figure}[!ht]
  \centering
    \includegraphics[width=0.8\textwidth]{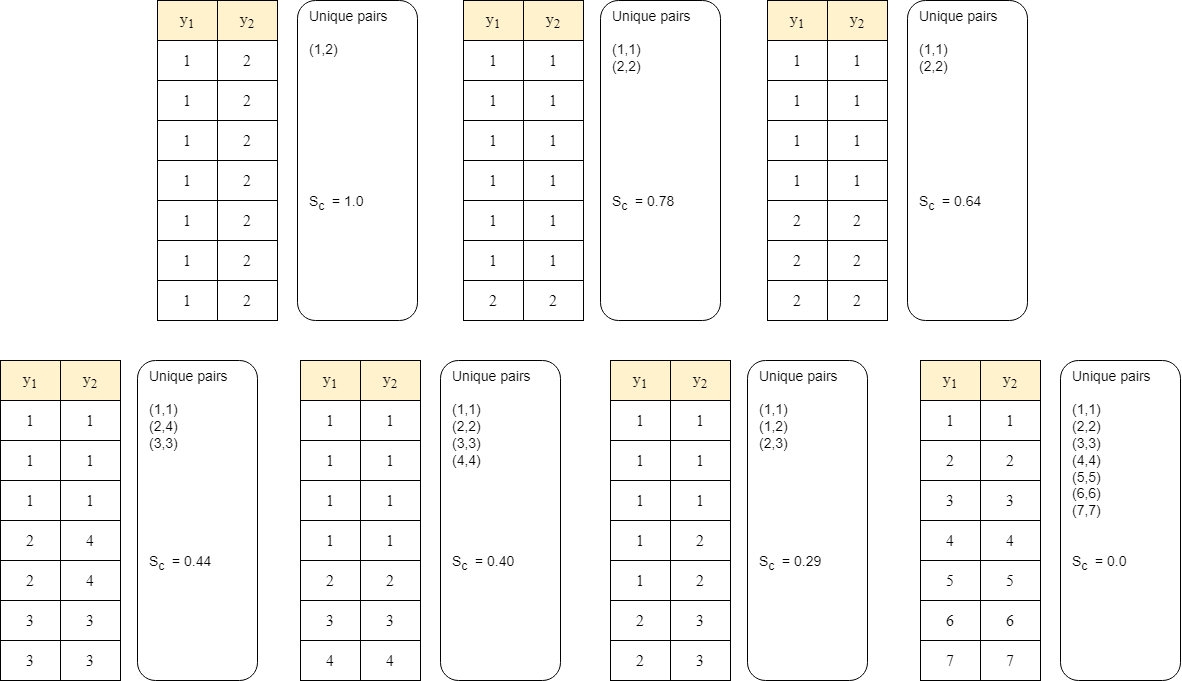}
    \caption{Seven examples of applying the proposed similarity correlation metric.}
    \label{similarity_example}
\end{figure}

As such, we propose the following two similarity metrics, $Sim_{Co}$ and $Sim_{Or}$, to fully capture the likelihood of two variables forming patterns. The $Sim_{Co}$ similarity metric can be used to determine the likelihood of two variables forming a constant, additive, or multiplicative pattern (both additive and multiplicative coherence can be viewed as constant coherence through vector manipulation \cite{henriques2014bicpam}):

\begin{equation}
\begin{split}
    Sim_{Co}(y_1,y_2) = \frac{\sum_{j=1}^{M} \frac{P(par_j)}{max(P(par_{j_{y_1}}),P(par_{j_{y_2}}))}}{M} \times \\ \times  (1 - \frac{\sum_{j=1}^{M}P(par_j) \times \log(P(par_j))}{\sum_{i=1}^{N} \frac{1}{N} \times \log(\frac{1}{N})}),
\end{split}
\end{equation}

\noindent where $P(par_j)$ represents a unique pair of values formed by the considering the values $P(par_{j_{y_1}})$ and $P(par_{j_{y_2}})$ extracted from the observation $j$ from vector $y_1$, and $y_2$, $M$ represents the number of unique $P(par_j)$. $M$ represents the number of observations in $y_1$ (or $y_2$). The formula captures two innate properties when forming patterns. The first property captured by:

\begin{equation}
    \frac{\sum_{j=1}^{M} \frac{P(par_j)}{max(P(par_{j_{y_1}}),P(par_{j_{y_2}}))}}{M},
\end{equation}

\noindent defines to the number of times a value in variable $y_1$ (or $y_2$) is matched with a different value from variable $y_1$ (or $y_2$). The second property captured by:

\begin{equation}
    (1 - \frac{\sum_{j=1}^{J}P(par_j) \times \log(P(par_j))}{\sum_{i=1}^{N} \frac{1}{N} \times \log(\frac{1}{N})}),
\end{equation}

\noindent is a proxy to the number of different patterns formed. The bigger the pattern, the more likely it is to have statistical significance \cite{henriques2018bsig}. To further clarify the desired matching between the variables, various examples, with different variables $y_1$ and $y_2$, are illustrated in Figure \ref{similarity_example}.

Although the $Sim_{Co}$ similarity metric accommodates patterns exhibiting constant, additive, and multiplicative coherence, it does not accommodate order-preserving properties. As such, we propose $Sim_{Or}$ to accommodate this type of coherence, it is defined as:

\begin{equation}
    S_{Or}(y_1,y_2) = \frac{\max\{\sum_{i=1}^{N}(y_{1_i} > y_{2_i}), \sum_{i=1}^{N}(y_{1_i} < y_{2_i})\}}{N},
\end{equation}

\noindent where $y_{1_i}$ (or $y_{2_i}$) corresponds to the value of observation $i$ in variable $y_{1}$ (or $y_{2}$), and $N$ to the number of observations in variable $y_1$/$y_2$.

\subsection{Biclustering and Merging}
Given a minimum support, $n_s$, biclustering is applied in the context of each partition to guarantee that no relevant partition-specific patterns are lost. After applying a biclustering algorithm to each partition, the extracted patterns are merged. To accomplish this, we implemented an efficient exhaustive procedure (Algorithm \ref{pattern_merge_algorithm}). The algorithm is divided into two parts. Part 1 joins patterns above a $n_s$ threshold, whilst part 2 focuses on the redundancy of the patterns, adding patterns to the final set of patterns only if the enclosed biclusters are maximal. The resulting set of patterns will contain all the patterns that could have originally been extracted from the original dataset, and all of them will be closed.

\subsection{Experimental setup}

To evaluate our solution, we considered patterns with constant coherence and the apriori algorithm, to extract the patterns. The upper limit of each partition is set to a predefined number of variables. An iteration consists of ten executions where a set of randomly selected variables is used, and the initial set consists of 4 variables. We gradually increment the number of variables for each iteration. Finally, to compare our results against alternative partitioning strategies, we considered three possible scenarios: for each set of randomly selected variables chosen, we 1) partition variables based on their likelihood to form patterns (local similarity approach), 2) partition based on dissimilarity amongst variables, and 3) randomly assigned each variable to a partition. Finally, we apply a simple one-sided t-Test to detect if there were any statistically significant improvements or deterioration in the execution times. This approach preserves simplicity by using an algorithm currently used in state-of-the-art approaches \cite{henriques2017bicpams}, while still assessing proof of concept.

\begin{algorithm}[!ht]
\scriptsize
\SetAlgoLined
\KwIn{clusters, nr\_obs, min\_sup}
\KwOut{list of patterns}
 final\_set $\gets$ []\;
 i $\gets$ 0\;
 \While{i $<$ size(clusters)} {
 ================ Part 1 ================ \\
    current\_set\_of\_patterns $\gets$ clusters[i]\;
    j $\gets$ i + 1\;
    \While{j $<$ size(clusters)}{
        \For{pattern in current\_set\_of\_patterns}{
            \For{pattern\_j in clusters[j]}{
                obs $\gets$ intersect\_observations(pattern, pattern\_j)\;
                \If{size(obs) / nr\_obs $>$ min\_sup}{
                    current\_set\_of\_patterns.append(obs, combine\_variables(pattern, pattern\_j))\;
                }
            }
        }
        j $\gets$ j + 1\;
    }
    ================ Part 2 ================ \\
    \For{p1 in current\_set\_of\_patterns}{
        added $\gets$ True\;
        \For{p2 in current\_set\_of\_patterns}{
            \If{\textbf{not} closed\_pattern(p1, p2)}{
                added $\gets$ False\;
            }
        }
        \For{p\_final\_set in final\_set}{
            \If{\textbf{not} closed\_pattern(p1, p\_final\_set)}{
                added $\gets$ False\;
            }
        }
        \If{added}{final\_set.append(p)\;}
    }
    i $\gets$ i + 1\;
 }
 
 \Return statistics\;
 \caption{Pattern Merge}
 \label{pattern_merge_algorithm}
\end{algorithm}

Given the aforementioned experimental conditions, we considered five datasets 

\noindent (\href{http://fimi.uantwerpen.be/data/}{http://fimi.uantwerpen.be/data/}), to test the scalability of our solution (see Table \ref{datasets}). The minimum support property in Table \ref{datasets} is inferred via the statistical significance test proposed by Henriques and Madeira \cite{henriques2018bsig}. Given a $\alpha = 0.05$ significance threshold, we assume that variables are uniformly distributed to compute the minimum support that guarantees the statistical significance of a pattern with minimum length 4. In Table \ref{datasets}, we can see that a high number of missing values highly impacted the minimum support, with the highest being 5\%.

\begin{table}[!ht]
\caption{Datasets used to test the scalability of the proposed solution and their respective properties: number of observations, number of variables, percentage of missing values (NaN (\%)) and minimum support needed for a pattern with four variables to have statistical significance (mininum support (\%)).}
\centering
\begin{tabular}{c|cccc}
  & nr. observations & nr. variables & NaN (\%) & min. sup. (\%)\\ \hline
 Accident & 340,183 & 468 & 93 & $>$ 0\\
 Chess & 3,196 & 75 & 50 & $>$ 5 \\
 Connect & 67,557 & 129 & 77 & $>$ 1\\
 Mushroom & 8,124 & 119 & 98  & $>$ 0\\
 Pumsp & 49,046 & 2,113 & 96 & $>$ 0\\
\end{tabular}
\label{datasets}
\end{table}

\section{Results and Discussion}

In order to evaluate our proposed approach, we used five datasets taken from the literature (\emph{Accident}, \emph{Chess}, \emph{Connect}, \emph{Mushroom} and \emph{Pumsp}). Table \ref{nr_patterns} includes the average number of patterns discovered in each iteration. Figure \ref{execution_time} shows (from images left to right) the execution time of the whole pipeline and merge algorithm, as well as the average number of observations and variables of the patterns extracted before the merging. Table \ref{p_values}, presents the \textit{p-value} between the execution times of the different approaches.

In Table \ref{nr_patterns}, the results reflect the impact of having a high number of missing values. The average number of patterns extracted in most of the datasets 
is considerably lower when compared to the potential maximum, or local maximum (\emph{Chess}), despite a low support threshold. This also translates into lower execution times along the merging stage for datasets where the number of patterns extracted is low, as seen in Figure \ref{execution_time}. 

In Figure \ref{execution_time}, when comparing the different approaches, results do not show a clear advantage in grouping variables based on similarity versus dissimilarity or randomness. For example, in the \emph{Accident} and \emph{Connect} datasets, we can see that the merging takes less time when variables are grouped based on similarity, but the overall execution time is not vastly affected by it. In the rest of the datasets (\emph{Chess}, \emph{Mushroom} and \emph{Pumsp}), the merging step has a higher execution time and it does slightly affect the overall execution time. Table \ref{p_values} complements the aforesaid, yet gives a more objective analysis of the improvements and deterioration of execution times. Both \emph{Accident} and \emph{Connect} have statistically significant improvements in the whole execution time using the local similarity approach (against both dissimilarity and random approaches), and no dataset yields significantly worse performance when using the similarity approach versus other approaches.

The explanation for the aforementioned execution times lies in the last two images at each row in Figure \ref{execution_time}. Overall, we observe that the average number of variables per pattern before merging is always higher in the similarity-based approach. This confirms our hypothesis that grouping similar variables creates patterns containing more variables than if we group variables by dissimilarity or randomly. The average number of observations per pattern is also lower in the similarity-based approach. This is to be expected since patterns with more variables have the same or a lower number of observations than patterns with a single variable. Given the preceding, the difference in execution times is in the little differences between the average number of variables and the average number of observations. Consider the results for the \emph{Chess}, \emph{Mushroom}, and \emph{Pumsp} datasets at 20 variables. The average number of variables is higher in the similarity-based approach, but the average number of observations is not considerably lower than the rest of the datasets. When the merging happens, the lists of pattern observations joined are bigger, translating into higher execution times in the merging operation.

\clearpage

\begin{table}[!ht]
\caption{The average number of patterns extracted ($\pm$ standard deviation) at the end of each iteration.}
\centering
\begin{tabular}{c|ccccc|}
\cline{2-6}
 & \multicolumn{5}{c|}{Nr. Variables} \\ \cline{2-6}
  & 4 & 8 & 12 & 16 & 20 \\ \hline
 \multicolumn{1}{|c|}{Accident} & $6\pm 3$  & $15\pm 7$  & $60\pm 39$ & $101 \pm 71$ & $145 \pm 65$ \\
 \multicolumn{1}{|c|}{Chess} & $8 \pm 2$ & $49 \pm 21$ & $403 \pm 462$  & $869 \pm 768$ & $ 6159 \pm 5252$  \\
 \multicolumn{1}{|c|}{Connect} & $7 \pm 2$ & $38 \pm 24$ & $108 \pm 48$ & $304 \pm 248$ & $612 \pm 435$ \\
 \multicolumn{1}{|c|}{Mushroom} & $6 \pm 2$ & $18 \pm 10$  & $62 \pm 23$ & $225 \pm 289$ & $218 \pm 104$ \\
 \multicolumn{1}{|c|}{Pumsp} & $5 \pm 1$ & $27 \pm 5$ & $69 \pm 21$ & $162 \pm 43$ & $514 \pm 319$\\ \hline
 \multicolumn{1}{|c|}{Maximum Nr. Patterns} & $15$ & $255$ & $4095$ & $65535$ & $1048575$\\ \hline
\end{tabular}
\label{nr_patterns}
\end{table}

\begin{table}[!ht]
\caption{One-tailed t-test with the null hypothesis that two related execution times have identical average (expected) values. In this case, p-values below 0.05 (marked as bold in the Table) reject the null hypothesis meaning that the mean of the similarity execution is \textbf{less} (or \textbf{greater}) than the mean of the dissimilarity/random execution.}
\centering
\begin{tabular}{cc|cc|cc|}
\cline{3-6}
  & & \multicolumn{2}{|c|}{Less*} & \multicolumn{2}{|c|}{Greater**} \\
\cline{3-6}
  & & Merge & Pipeline & Merge & Pipeline \\ \hline
 \multicolumn{1}{|c|}{\multirow{5}{*}{Simimilarity/Dissimilary}} & Accident & 0.08 & 0.11 & 0.91 & 0.88\\
 \multicolumn{1}{|c|}{} & Chess & 0.81 & 0.81 & 0.18 & 0.18 \\
 \multicolumn{1}{|c|}{} & Connect & 0.14 & \textbf{0.01} & 0.85 & 0.98 \\
 \multicolumn{1}{|c|}{} & Mushroom & 0.92 & 0.79 & 0.07 & 0.2 \\
 \multicolumn{1}{|c|}{} & Pumsp & 0.81 & 0.80 & 0.18 & 0.19 \\ \hline
 \multicolumn{1}{|c|}{\multirow{5}{*}{Simimilarity/Random}} & Accident & \textbf{0.03} & \textbf{0.04} & 0.96 & 0.95\\
 \multicolumn{1}{|c|}{} & Chess & 0.81 & 0.79 & 0.18 & 0.20\\
 \multicolumn{1}{|c|}{} & Connect & 0.10 & \textbf{0.01} & 0.89 & 0.98 \\
 \multicolumn{1}{|c|}{} & Mushroom & 0.88 & 0.82 & 0.11 & 0.17 \\
 \multicolumn{1}{|c|}{} & Pumsp & 0.81 & 0.80 & 0.18 & 0.19 \\ \hline
 \multicolumn{6}{l}{* The average execution time of the Similarity approach is less than the} \\
 \multicolumn{6}{l}{Dissimilarity/Random} \\
 \multicolumn{6}{l}{** The average execution time of the Similarity approach is greater than the} \\
 \multicolumn{6}{l}{Dissimilarity/Random}
\end{tabular}
\label{p_values}
\end{table}

\begin{adjustwidth}{-2cm}{1cm}
\vspace*{-3.4cm}
\begin{minipage}{1.25\textwidth}
\centering

\textbf{\large Accident} \\

\includegraphics[height=1.4in]{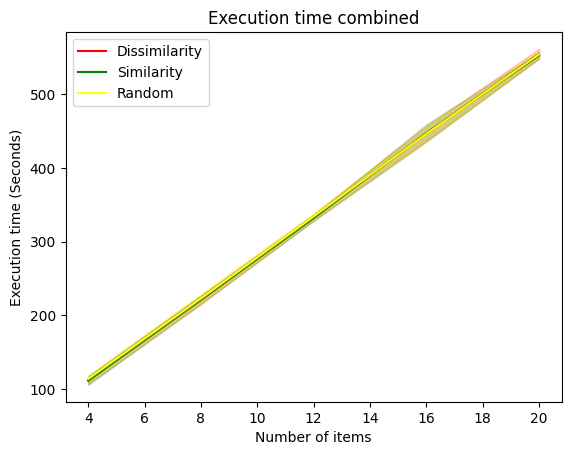}
\includegraphics[height=1.4in]{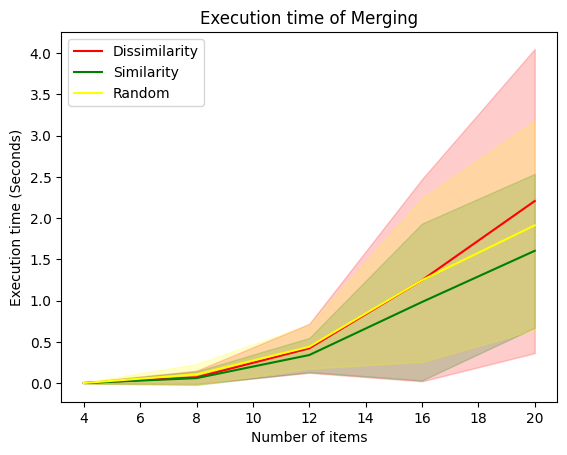}
\includegraphics[height=1.4in]{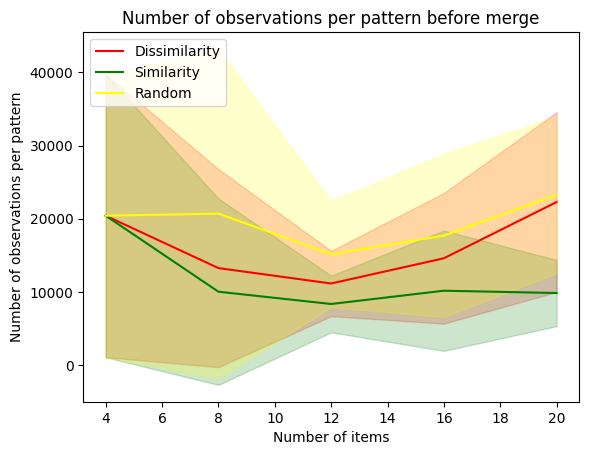}
\includegraphics[height=1.4in]{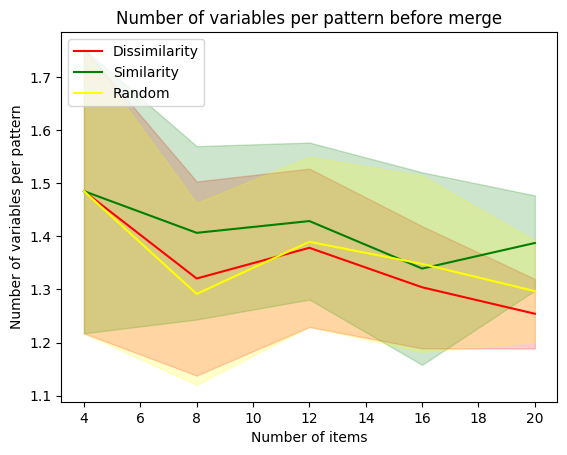}

\textbf{\large Chess} \\

\includegraphics[height=1.4in]{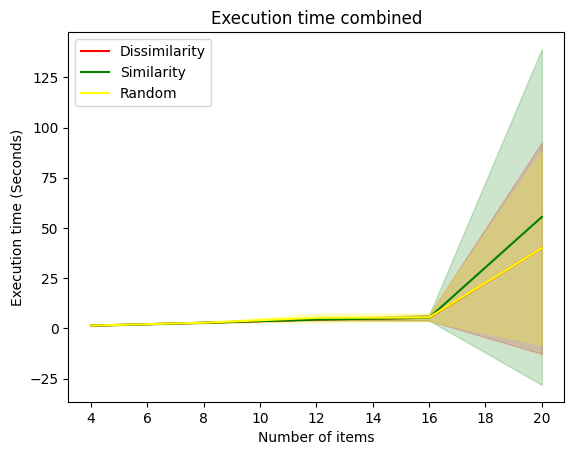}
\includegraphics[height=1.4in]{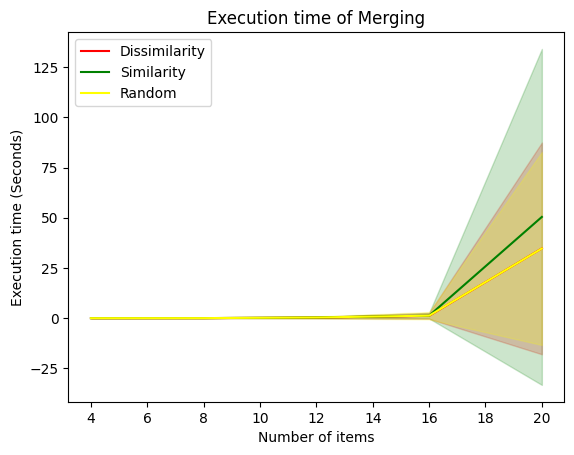}
\includegraphics[height=1.4in]{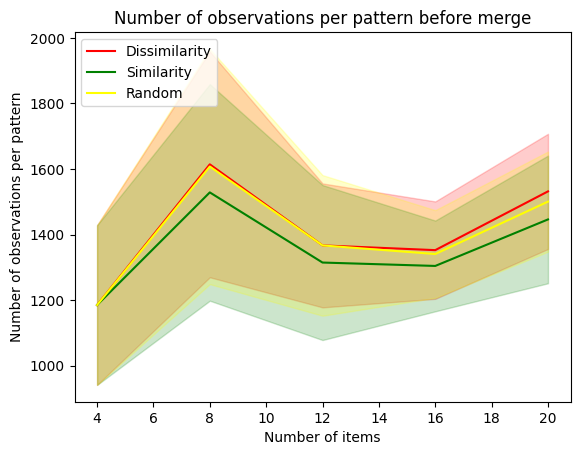}
\includegraphics[height=1.4in]{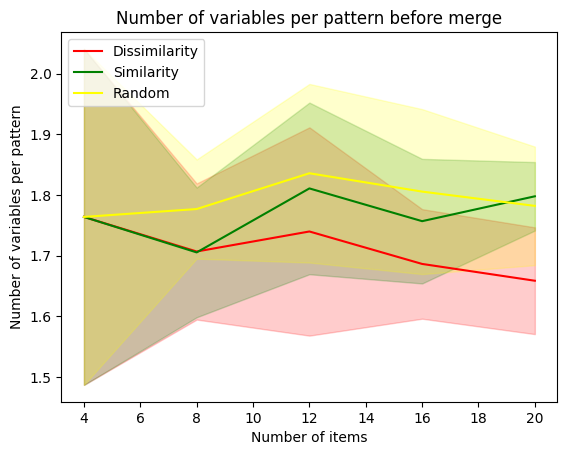}

\textbf{\large Connect} \\

\includegraphics[height=1.4in]{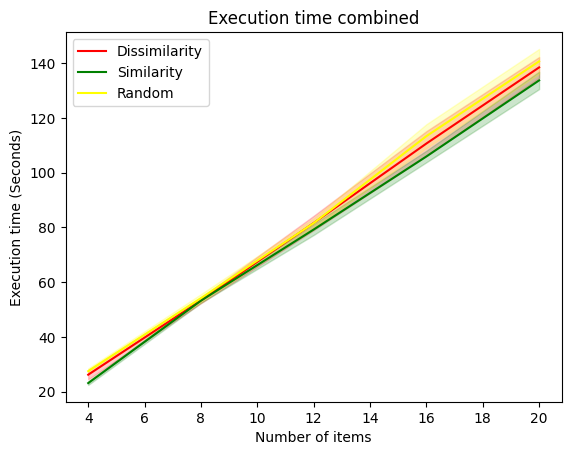}
\includegraphics[height=1.4in]{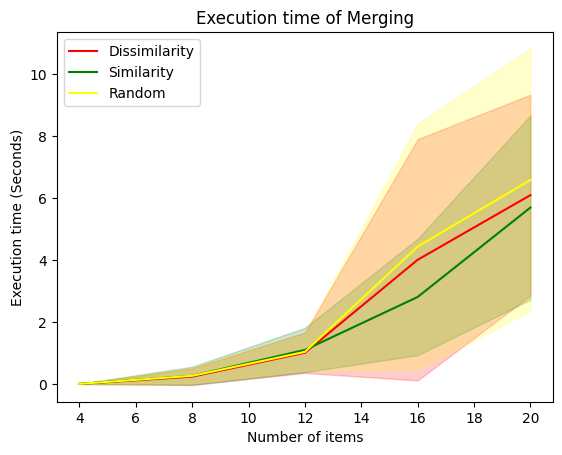}
\includegraphics[height=1.4in]{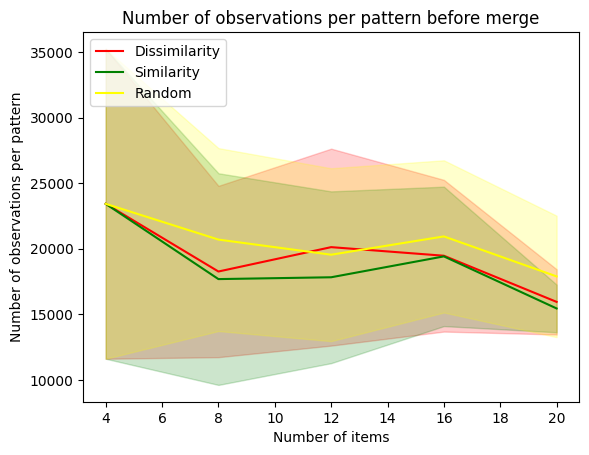}
\includegraphics[height=1.4in]{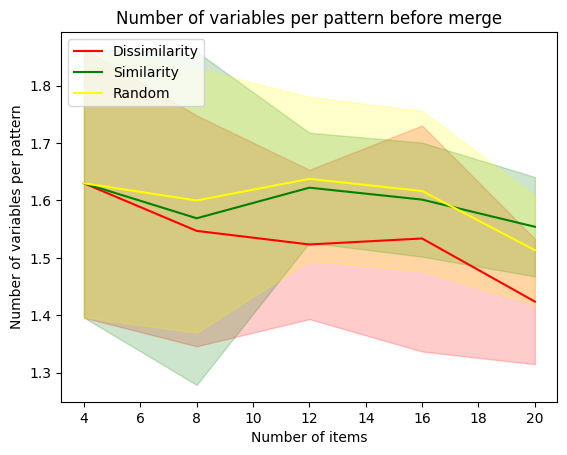}

\textbf{\large Mushroom} \\

\includegraphics[height=1.4in]{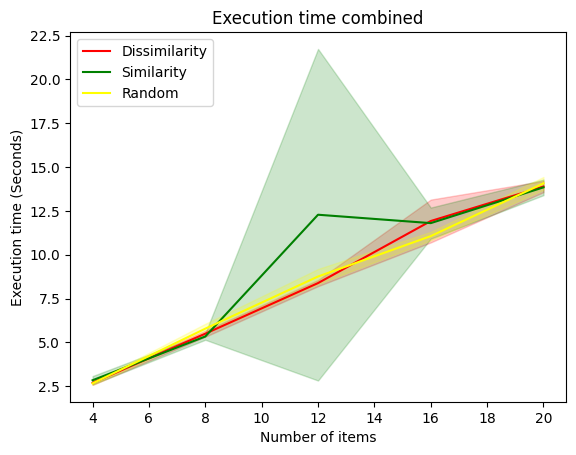}
\includegraphics[height=1.4in]{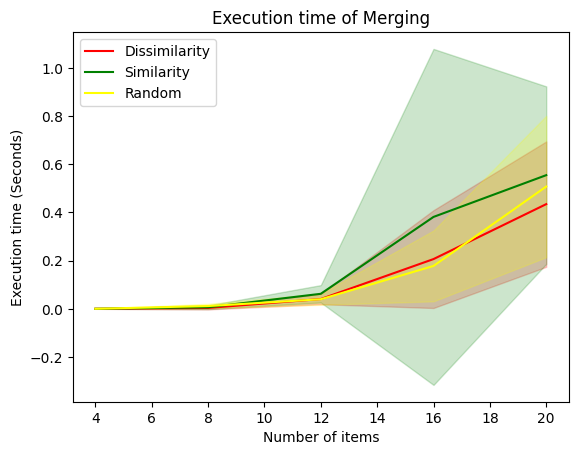}
\includegraphics[height=1.4in]{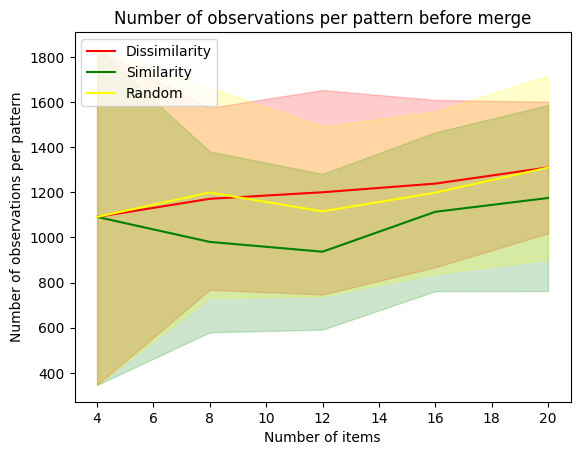}
\includegraphics[height=1.4in]{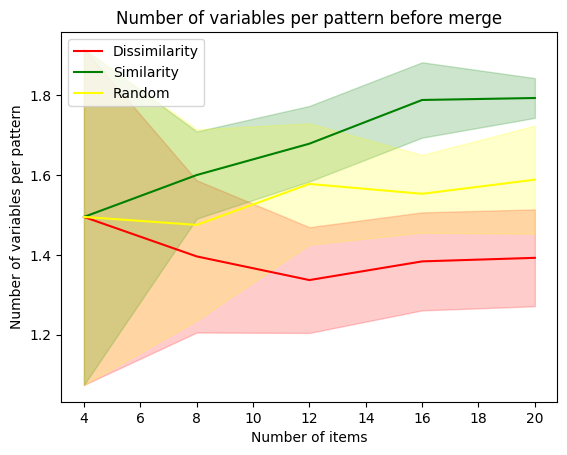}

\textbf{\large Pumsp} \\

\includegraphics[height=1.4in]{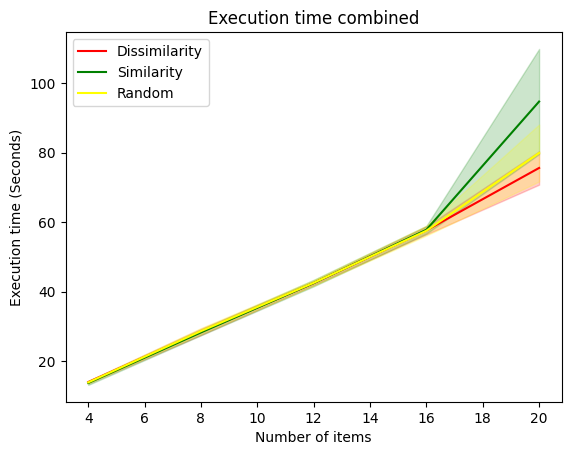}
\includegraphics[height=1.4in]{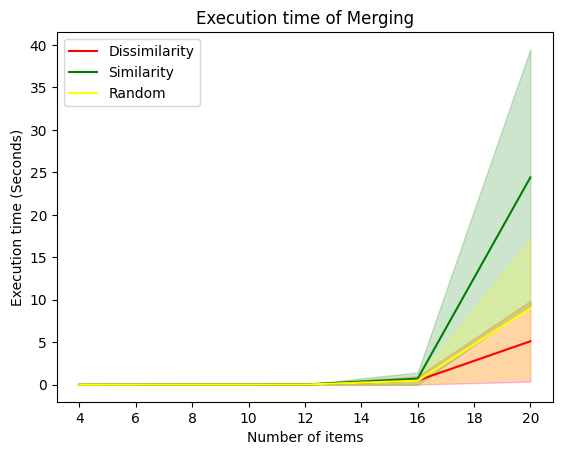}
\includegraphics[height=1.4in]{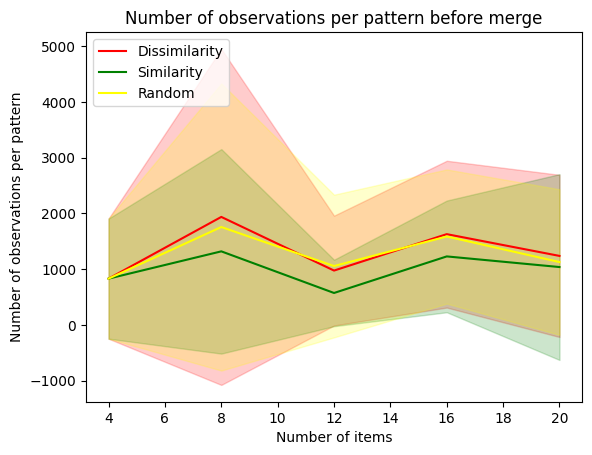}
\includegraphics[height=1.4in]{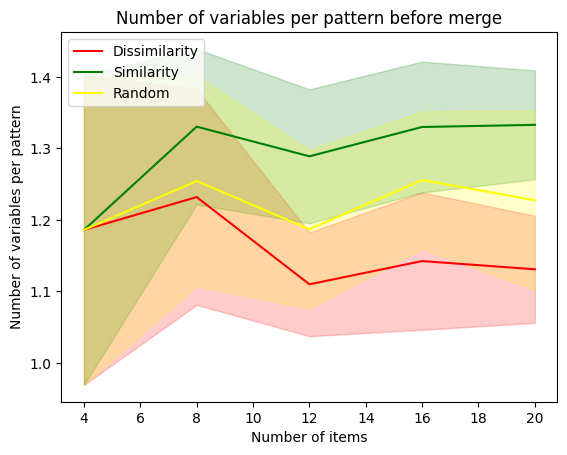}

\captionof{figure}{A set of four images per dataset. In each of the dataset (\emph{Accident, Chess, Connect, Mushroom, Pumsp}) a set of four images is presented. From left to right in each set of images they represents: 1) the average execution time of the whole pipeline, 2) the average execution time of the merge algorithm, 3) the average number of observations of the patterns extracted before the merging, and, 4) the average number of variables of the patterns extracted before the merging. Each image also contains three lines, green (similarity based), yellow (randomly picked), red (dissimilarity based), that correspond to the grouping technique applied. Finally, the corresponding color area represents the standard deviation to the averages in each image.}
\label{execution_time}
\end{minipage}

\end{adjustwidth}

\section{Conclusion}

This work proposed a new approach to scale biclustering algorithms by partitioning the dataset into non-overlapping clusters based on variable similarity with the premise that patterns are well-formed per partition, and won't need to entail considerable expensive merging steps. The execution times achieved by this approach in some datasets proved to be statistically significantly lower when compared with partitions based on variable dissimilarity or randomly clustered. Although some results are promising, more work is required to estimate the potential of this solution based on: 1) data properties, 2) the type of patterns extracted, and 3) the biclustering algorithm used. Further development can still be done regarding the merging principles.




\end{document}